# A versatile metasurface enabling superwettability for self-cleaning and dynamic color response

Jinlong Lu, Basudeb Sain, Philip Georgi, Maximilian Protte, Tim Bartley, and Thomas Zentgraf[*]

*Paderborn University, Department of Physics, Warburger Str. 100, 33098 Paderborn, Germany*
[*]*Corresponding author. Email: thomas.zentgraf@uni-paderborn.de*

**Abstract**

Metasurfaces provide applications for a variety of flat elements and devices due to the ability to modulate light with subwavelength structures. The working principle meanwhile gives rise to the crucial problem and challenge to protect the metasurface from dust or clean the unavoidable contaminants during daily usage. Here, taking advantage of the intelligent bioinspired surfaces which exhibit self-cleaning properties, we show a versatile dielectric metasurface benefiting from the obtained superhydrophilic or quasi-superhydrophobic states. The design is realized by embedding the metasurface inside a large area of wettability supporting structures, which is highly efficient in fabrication, and achieves both optical and wettability functionality at the same time. The superhydrophilic state enables an enhanced optical response with water, while the quasi-superhydrophobic state imparts the fragile antennas an ability to self-clean dust contamination. Furthermore, the metasurface can be easily switched and repeated between these two wettability or functional states by appropriate treatments in a repeatable way, without degrading the optical performance. The proposed design strategy will bring new opportunities to smart metasurfaces with improved optical performance, versatility and physical stability.

## 1. Introduction

Unlike conventional optical components based on light propagation and refraction in bulk materials, metasurfaces employ specially distributed nanoantennas to tailor light with desired amplitude, phase, and polarization.[1-4] A broad range of functionality based on the ability of nanoantennas to accurately modulate light have been experimentally achieved, such as structural color,[5-10] metaholograms,[11-16] metalenses,[17, 18] and many others.[19, 20] The nanoscale tailoring of light inspires improvements to exciting optical functionality and expands the potential applications of metasurfaces in ever more directions. However, the fragile nanostructures are typically highly sensitive to dust and other contaminants, which ultimately limits their wider application. Until now, there is no solution to efficiently clean or protect from the dust contamination on a metasurface that consists of specifically designed antennas. Hence, finding a feasible strategy to clean the metasurface without affecting its optical performance or damage the nanostructures is extremely promising for various practical applications. Furthermore, future development of metasurfaces entails not only continuous efforts to improve the optical performance, but also accessible strategies to enhance the device adaptability and effectiveness in complex application conditions.[21-23] This means metasurface with enhanced optical performance, physical stability and smarter functionality are important properties that need to be considered when one designs the device.

Similarly to metasurfaces, bio-inspired surfaces comprising micro/nanosized structures show enhanced physical characteristics. One important example is superwettability, i.e., a superhydrophilic surface with a water contact angle (CA) < 10° or a superhydrophobic surface with CA > 150°. They tend to be more robust and smarter to a variety of harsh environments,[24] with versatile applications like surface self-cleaning,[25, 26] anti-fog/corrosion,[27] water separation/control can be found.[28-30] Recently, seminal works show several advantages to combine the devices' optical performance with the wettability design.[31-37] Standing out among them, an ultrafast humidity-responsive structural color was realized by taking advantage of the hydrophilic nanoporous $TiO_2$.[36] A self-cleaning antireflective surface was shown with random nanopillars resulting in the superhydrophobic state, which is based on the well-known lotus leaf effect to effectively roll off the water droplet and contamination.[35] A water-selective metasurface with hydrophilic/hydrophobic parts is used to actively modulate and steer the reflected light between the two states of wetting and drying, which shows the possibility to enrich the functionality.[37] Insights of successfully including wettability design could improve the optical performance of metasurfaces, and at the same time equip them with smart self-cleaning or more functionalities. However, the wettability of a surface, typically expresses along with a large area of rougher micro/nanostructures, with structural requirements that are contrary to the design of optical metasurfaces. This makes the achievement of superwettability on the intentionally arranged nanoantennas challenging and raises the question: Is it possible to realize desired wettability on optical metasurfaces regardless of the small area and specific antenna distribution? Will it provide us another strategy to achieve better optical performance or more functionalities for metasurfaces?

As one of the specific applications, a structural color metasurface can generate a vivid image via a spatial arrangement of resonant nanostructures,[5, 7-10] whereas a dynamic color response can be efficiently achieved by changing the refractive index that surrounds the antennas. These properties make it a suitable candidate to directly investigate the optical and self-cleaning performance of the metasurface.[32, 36] Here, we design a structural color metasurface that is embedded in a large area wettability supporting structures, which enable the possibility to achieve two superwettability states by proper treatments. These wettability states with efficient color response or self-cleaning ability can be easily switched between each other, and no obvious degradation of the optical performance is observed during several repeatability tests. The strategy to use a wettability design provides us an economic solution to achieve versatile metasurfaces with an improved optical performance and physical stability.

## 2. Results and Discussion
### 2.1. Working principle and fabrication of the metasurface

The proposed amophous Si (a-Si) metasurface possesses two superwettability states, which can be easily switched between each other by the introduced hydrophilic or hydrophobic treatment. Each wettability state is linked with a unique functionality as shown in **Figure** 1a. The quasi-superhydrophobic state (CA > 130°): the large CA causes the droplets on the surface to easily slide off, which can be used to efficiently self-clean the tiny particles on the metasurface without destroying the nanoantennas. The superhydrophilic state (CA < 10°): A water droplet on the surface spreads very fast and efficiently forms a flat water film that perfectly immerses the antennas inside. This state provides the possibility to design a metasurface with a dynamic color response (due to the change of surrounding environment), which is further improved due to the spreading behavior of water. More important, since the wettability of a surface is determined by the surface roughness and free energy together, it is possible to achieve a wettability transition between different states, just by modifying the surface free energy while keeping the surface structures unchanged.[38-40] The possibility to achieve a wettability transition on the metasurface is important here, which would guarantee the realization of more functionalities.

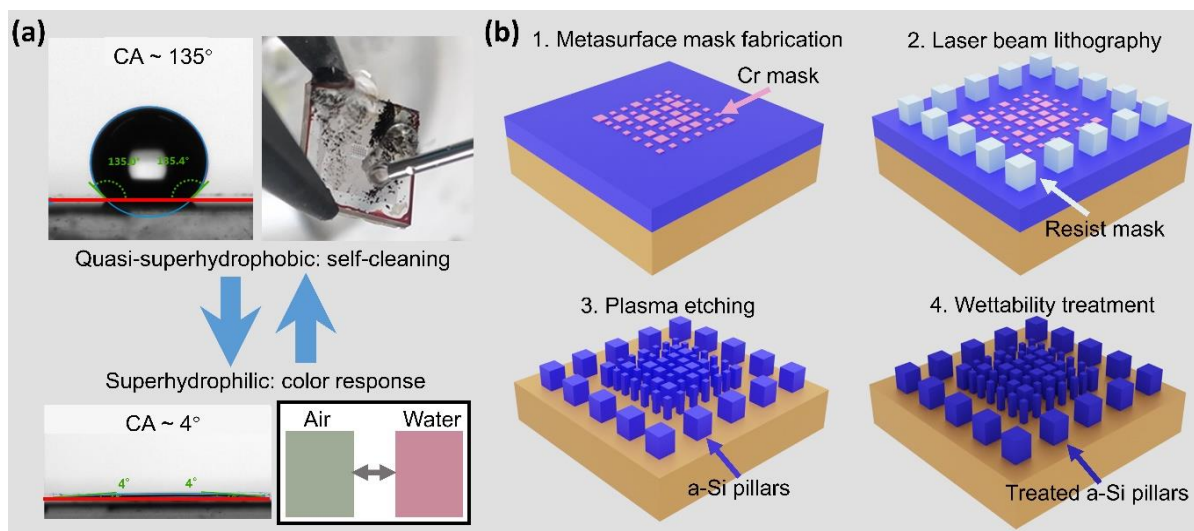

**Figure 1.** Diagram of functionality or wettability transition, and illustration of sample fabrication. (a) Two superwettability states can be switched to each other with Octadecylphosphonic acid (ODP) modification (hydrophobic treatment) or $O_2$ plasma treatment (hydrophilic treatment), each state is linked with a unique performance of the metasurface: the quasi-superhydrophobic state can be used to self-clean the dust-contamination on the metasurface, the superhydrophilic state can be used to show the color change efficiently with water. (b) The sample is fabricated in four main steps: (Step 1) fabricating the Cr mask of optical metasurface by standard electronic beam lithography (EBL), (Step 2) fabricating the resist mask of wettability supporting structures by laser beam lithography, (Step 3) plasma etching and removing of masks, (Step 4) corresponding treatment to achieve the desired wettability state.

To fabricate a sample (constructed by the same material) with superwettability, the well-designed surface such that it increases surface area (Wenzel state) or facilitate entrapment of air under the droplet by reducing solid area fraction (Cassie state) would be helpful for better wettability performance (see *Supporting Information Section 1*). This means that a metasurface with specifically distributed nanoantennas cannot always be the suitable design to achieve the desired wettability. On the other hand, in almost all cases, it's unnecessary and too time-consuming to fabricate an optical metasurface with a large area. Here, the problem is solved by the following multi-step fabrication strategy, which provides both required wettability and optical functionality at the same time as shown in **Figure** 1b: first, the nanosized metasurface mask is fabricated by a standard electron beam lithography (EBL) process (Step 1); then, a following UV laser lithography is conducted to fabricate the resist mask of the structure in a large area, that is used to support the wettability (Step 2); further, the sample is etched with both masks together and followed by removing the residues (Step 3); finally, the corresponding wettability treatment is used to induce the superwettability state (Step 4). In detail: the wettability supporting structures are constructed by uniform arrays (a square pillar size around 1 μm, the distance between adjoin pillars is 2 μm) that can improve the wettability performance according to available published results,[26] which we confirmed in tests without the metasurface area. The etching step uses resist and Chromium (Cr) masks together, which provides protection against etching for both metasurface and wettability supporting structures at the same time. Wettability treatment by Octadecylphosphonic acid (ODP) modification results in a surface with quasi-superhydrophobic state, while a fast $O_2$ plasma treatment can change the surface back to the superhydrophilic state. The $O_2$ plasma treatment induces more polar bonds on the antenna surface, which makes it hydrophilic; while the ODP modification coats a low-surface-energy molecule on the surface and results in a non-polar and hydrophobic state.[27, 39]

Note that the sample still possesses a normal state that is without performing any wettability treatment, which shows a CA around 90°. The stored state after hydrophilic treatment is also included in this normal state due to the similar wettability behavior as shown in *Supporting Information Section 2*. This normal state impedes the self-cleaning process according to previous research, as a water droplet forms a curved shape on the surface and partly penetrates the volume between the antennas.[26] The problem arises when water droplets adhere to the surface tightly and most of the contamination cannot be washed away. On the other hand, the poor hydrophilic performance makes it difficult to observe the designed dynamic color response due to the air in the anttennas and light scattering influence. Taking water used in this work as an example, the curved shape of the droplet results that no color imprint image can be recorded; or the randomly induced air between the water film and nanoantennas generates colors that are out of control as shown in *Supporting Information Section 2 and Video 1*.

## 2.2. Optical and wettability characterizations of the metasurface

The rigorous coupled-wave analysis (RCWA) is used to optimize three unit cells for the metasurface with a large size difference, which perform notable color changes after changing the environment from air to water. The height of the a-Si antenna is $H = 700$ nm, which can be used to design metasurface with other optical functionalities. Square unit cells with a constant period of $P = 800$ nm and square antennas with sizes of 260 nm, 410 nm and 560 nm on an ITO coated substrate are chosen based on the simulation. The reflective spectra of uniformly arranged arrays in air and the corresponding color response (reflective spectra) by immersing the same structure in deionized water are shown in **Figure** 2. Here, the simulated spectra are calculated at normal incidence, the experimental results are recorded by a 10x objective lens with NA of 0.3, and the color (in RGB space, see calculation method in *Supporting Information Section 3*) of each spectrum is used to plot the data. Note that, as the reflection spectrum of the 260 nm antenna (**Figure** 2a) is quite weak compared to the other two sizes (especially in the range of 450 - 600 nm which is responsible for the visible color reception), we use two colors that are close to the real results to show the data. Besides, most of the observed colors and experimentally recorded spectra match with the simulation to some extent. For the 410 nm size antennas (**Figure** 2b), two obvious peaks around 500 - 550 nm and 650 - 700 nm are present in air, resulting in the observed green color; while the relative peak intensity around 500 - 550 nm drops a lot in water, with the other peak at long wavelength range shifts to around 650 nm. The spectral difference in air and water causes the observed green-red color change (inset color image in **Figure** 2b). For the 560 nm size antennas (**Figure** 2c), two peaks around 650 nm and 700 nm are present both in air and water, and an intensity drop for both of them is observed in water. The slight peak shift does not obviously change the color in air and water, while the larger color difference observed in experiments could be linked with the dropping off the intensity in water. Still, some differences can be found between the simulation and experiment, which is possibly caused by a combination of multiple influences including incident angle, higher-order diffraction due to the large period and fabrication tolerances (more details in *Supporting Information Section 4*). Here in the experiment, the illumination white light source is focused by an objective lens, which results in a larger range of incident angles compared to the normal incident case, while the fabrication tolerances always result with a non-uniform rectangular solid antenna. Therefore, the simulation results, with only one specific case of antenna geometry is calculated, just show some correspondence with the experiments. The influence from fabrication may even obvious when the antenna size is 260 nm, which is highly possible for a higher sensitive to the size and the low reflection of this design. Metasurfaces showing QR code images that are made of these three antennas are used to visualize the dynamic and self-cleaning property in the later part. Recording the QR code image directly shows the metasurface performance at different steps.

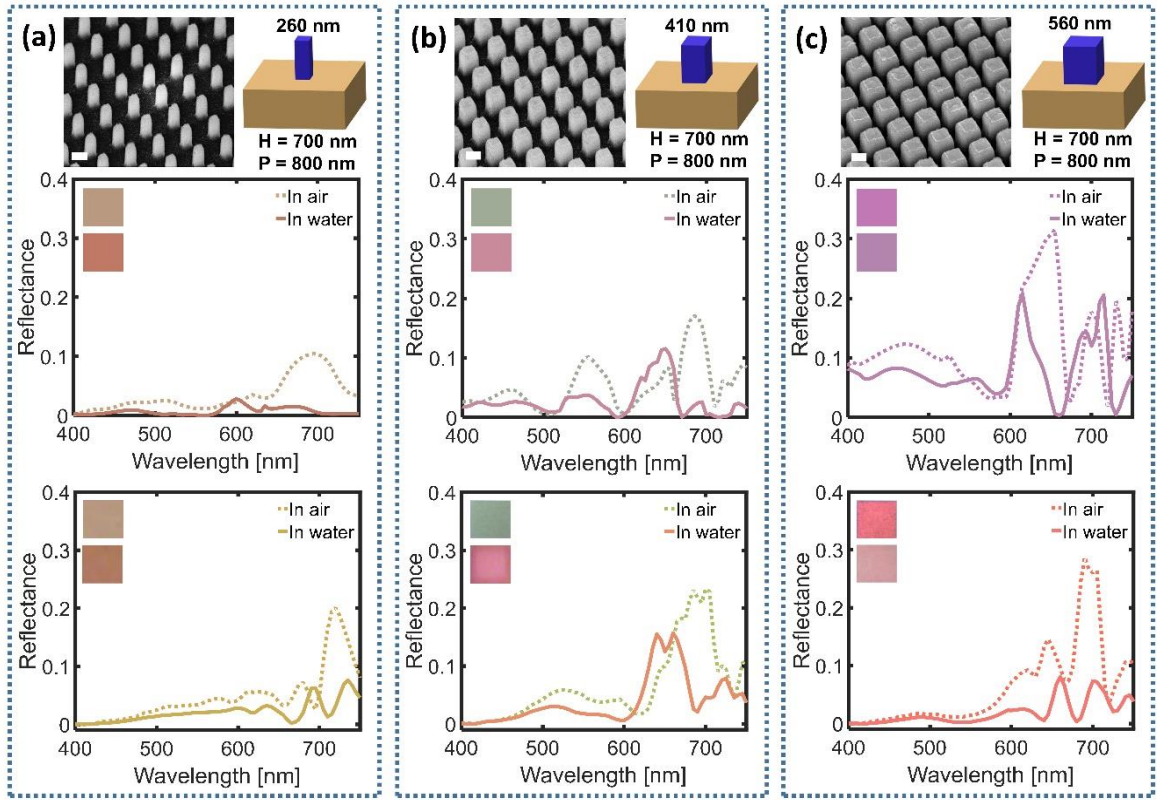

**Figure 2.** Reflective spectra from simulations and measurements for uniform arrays with three different antenna sizes in air or water. (a) 260 nm, (b) 410 nm, (c) 560 nm. Top row: SEM images and illustration of different antennas. Middle row: simulated spectra with inset images showing the color in air (top) or water (bottom). Last row: experimentally recorded spectra and color images in air (top) or water (bottom). The corresponding RGB color (except simulation in (a)) of each spectrum is used to plot the data. Scale bars are 400 nm.

Next, CA of two samples (with and without wettability supporting structure) with the QR image pattern at different states are shown in **Figure** 3. The SEM images are shown in the left column, with three recognizable areas in the QR code image area corresponding to the three antenna sizes. Each metasurface with the QR code image has an area of 180 x 180 μm$^2$ with a total number of 225 x 225 pixels. After fabrication without any further treatment, CAs around 70 - 100° are measured, which is a moderate value that belongs to the normal state. On the other hand, both wettability treatments of the sample without supporting structures do not change the CA remarkably (bottom row, CA ~26° to 108° after hydrophilic and hydrophobic treatment). In stark contrast, the sample with wettability supporting structures achieves a CA less than 10° after the hydrophilic treatment, while this value boosts to larger than 130° after the hydrophobic treatment.

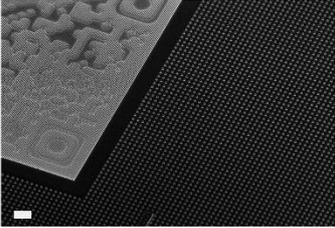

**Figure 3.** CA evolution of two samples with and without wettability supporting structure. The uniform area in the SEM image (top left corner) corresponds to the wettability supporting structures on the sample. Both samples do not show extraordinary wettability at the normal state (second column, CA ~ 101° and 73°). The sample with supporting structure shows a larger CA difference after the hydrophilic and hydrophobic treatment (last two lines, CA ~ 5° to 134° vs CA ~ 26° to 108°). Details of the CA measured results: the red lines show the baselines of the surface, the curved blue lines corresponding to the profile of the water droplet, the green lines show the tangent of the water droplet, the CA is the angle between the red and the green line in each image. The reported CA in the figure is an averaged value over three measurements. Scale bars are 10 μm.

The larger CA difference is what we desire as it may provide both better color response and self-cleaning property. Note here, the modification by the ODP molecules with lower-surface-energy only forms a very thin layer on the sample, this process almost does not influence the optical performance of the metasurface as shown by the structural color and SEM image in the *Supporting Information Section 5*. On the other hand, the design strategy by embedding the metasurface within the wettability supporting structures makes it possible to further improve the wettability performance through tuning the structure or modification method.

## 2.3. Dynamic color response and self-cleaning performance with water

As shown in **Figure** 4, under white light illumination, similar QR code color images can be recorded in air for both samples by different treatments (results of the sample without supporting structure are shown in *Supporting Information Section 6*). It's important to note that the wettability treatments by the ODP modification and $O_2$ plasma do not change the initial QR code color images in air, which is necessary for the optical performance of the metasurface. In stark contrast, only the state after hydrophilic treatment can be used to record the color change of the QR code image in water, the other states do not show any clear image due to the formation of a water droplet on the surface. Here, the unclear image is mainly caused by two threats and cannot be improved even with further focus tuning as shown in *Supporting Information Video1*. First, light is easily influenced by the curved surface of the water droplet on the metasurface; second, air is trapped between the antennas because of the poor hydrophilic performance, which further causes aberrations of images due to the randomly scattering of reflected light from the antennas. Both images in air or water are similar to the simulation (right column), especially the corresponding green-red color change from air to water. A video shows the dynamic color change as observed by a top-view commercial microscope (see *Supporting Information Video 1*), which proves the importance of the wettability design for the better dynamic response of the metasurface. Here, although these two samples after hydrophilic treatment show a similar static image, the better hydrophilic performance of the sample with wettability supporting structures is helpful for a more efficient response as shown by Mohd-Noor et.al., who designed an ultra-fast response humidity optical sensor with hydrophilic $TiO_2$.[36] Unfortunately, our current setup cannot be used to observe the entire dynamic behavior to compare more details of the response time. However, a spreading time of the water droplet on the superhydrophilic surface that is less than 20 ms can be used to estimate the dynamic response, which is a similar level to their measurement with the optical sensor.[36, 41] The difference of color change with different wettability is explained by an illustration of the water droplet state on each metasurface in **Figure** 4 bottom row.

Furthermore, the dynamic response of a metasurface by changing the environmental refractive index with water degenerates when the sample is stored after some time (see normal state). The degeneration of the superhydrophilic state is observed for different kinds of surfaces, which mainly results from the absorption of polar groups in air. With a proper wettability treatment or design, it is possible to keep this function for a longer time.[39, 40, 42] Here, the hydrophilic performance with the wettability supporting structures can keep the state around double time compared to the other samples. The improvement is highly possible for the better initial-wettability with the supporting structure.

Except for the optical response, the perfect matching between the surrounding liquid and the metasurface can also be useful for more specific applications like underwater detecting or imaging if we recall the refractive index matching method of traditional optical components.[43] A similar strategy can also be used to fabricate superoleophilic surfaces, which further extends the choice of matching solution.[27]

**Figure 4.** Color response of the metasurface with wettability supporting structures in air and water. All the states clearly show color images in the air (first line) similar to the simulation, while only the hydrophilic treated state (second column) show color change image with water droplet. The corresponding diagrams illustrate the water droplet state on the surface (last row), showing the importance of perfect matching between water and nanoantennas to the dynamic color response. Scale bars are 20 μm.

For the metasurface with self-cleaning property, we notice Wu et.al. recently demonstrated that antennas constructed by $TiO_2$ can be used to remove some chemical contamination, which is mainly due to the photocatalytic reaction.[23] Beside the chemical contamination, particle or dust adhere on the metasurface during daily usage is also extremely critical for most metasurface applications.[35] Based on our observation, once the particle sticks on the sample, especially for the small particles that can resident inside the gap between the antennas, it's almost impossible to remove them anymore by water cleaning (see *Supporting Information Section 7 and 8* for further or repeating cleaning). Due to the weakness of nanoantennas, it's also hard to clean the particles by other methods. Therefore, our design with supporting structures that can enable one step removal of particles is highly promising.

In the following, we test the self-cleaning property and compare the performance. The two states show better hydrophobic properties after the ODP modification is used, as a larger CA is helpful for the self-cleaning of particle-based contaminations.[26, 27, 44, 45] During cleaning, the sample is placed by a titled angle around 45°, with carbon particles as the artificial contamination to cover the whole sample area. Then, water at room temperature is dropped onto the metasurface to clean the surface. We observe the particles stick outside the droplets and are washed away from the metasurface area when the droplet is rolling over the contaminated area (cleaning process in *Video 2*). For the other sample without supporting structure and with the same cleaning condition, there is always a large water droplet that adheres to the surface together with lots of particles during the whole process, which is caused by the poor hydrophobic area outside the metasurface region.[27, 30, 44, 45] Therefore, the balance of capillary and adhesion force between the droplet and the contamination on the substrate that determines the friction force of drops during self-cleaning causes most particles always stuck there, and they cannot be cleaned away anymore.[44]

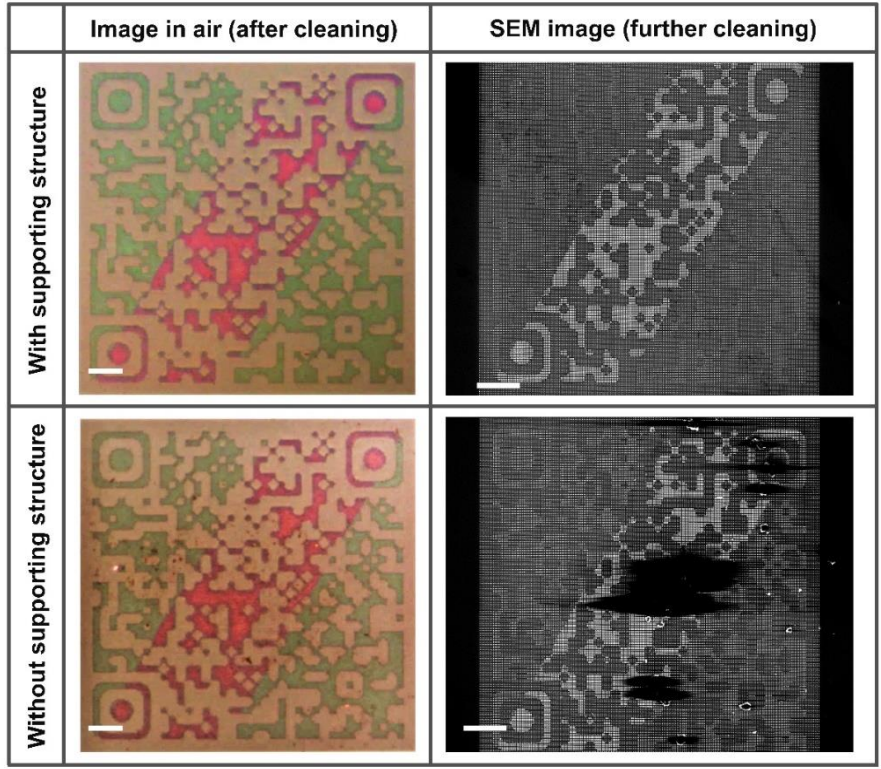

**Figure 5.** Comparison of the self-cleaning performance of metasurface. No particle is observed on the sample with wettability supporting structures (both in the optical and SEM image, top row). Lots of particles are present in the images from the sample without supporting structures (bottom row). Scale bars are 20 μm.

The QR code images after the self-cleaning process can be used to visualize the cleaning performance directly, with the same image as the initial one that is recorded for the metasurface embedded within the wettability supporting structures (**Figure** 5). However, lots of small dark areas caused by the contamination are found on the other sample without wettability supporting structure, and further processes cannot clean it anymore as shown in the images at the bottom (more details in *Supporting Information Section 7)*. Further check of a metasurface field after cleaning is performed by SEM, and we do not find any obvious particle left in the QR code image area for the sample with supporting structure, the few particles present at the metasurface edge or supporting structure is mainly caused by the blank area without any structures and imperfect arrangement of uniform arrays.[26] The obvious difference in cleanness shows the hydrophobic design with a large area improves the self-cleaning performance a lot.

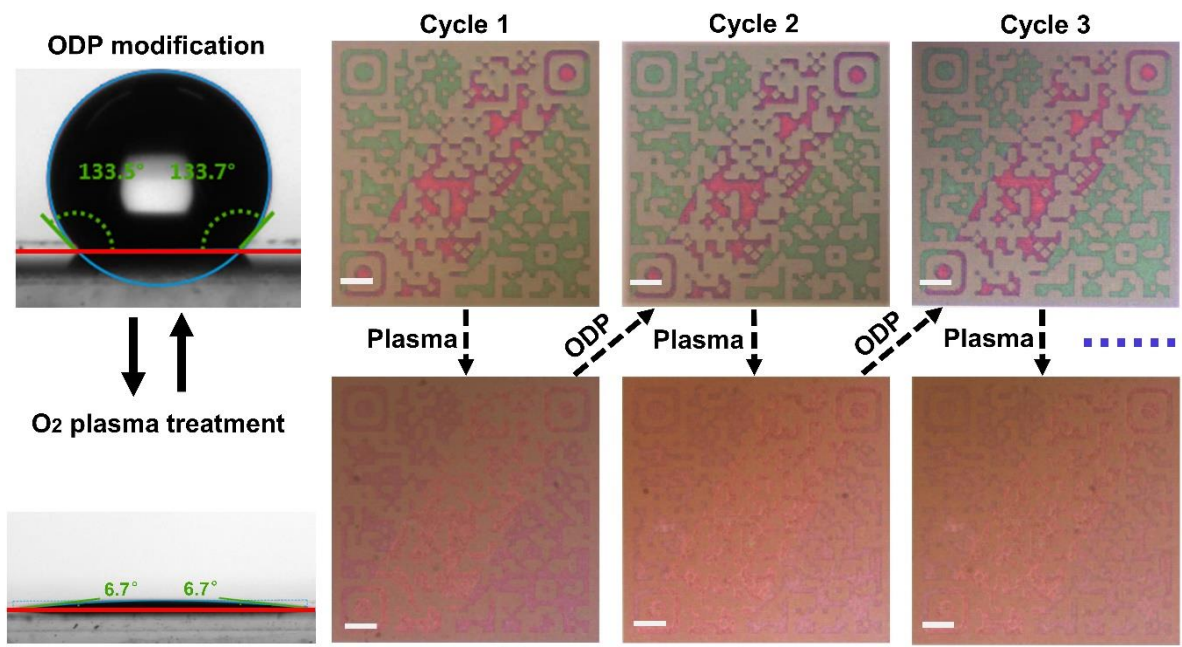

**Figure 6.** Repeatability testing for three times between the self-cleaning and dynamic response. The process starts with modifying the surface with ODP, followed by self-cleaning. Then the sample is treated by $O_2$ plasma, after which the dynamic response image in water is recorded. No noticeable degeneration of optical performance is observed after three cycles. Scale bars are 20 μm.

More interestingly, for the metasurface with supporting structure, we can randomly choose our metasurface at the self-cleaning state or at the color response state that is guaranteed by the corresponding wettability treatment. With $O_2$ plasma treatment or ODP modification, the transition between superhydrophilic and quasi-superhydrophobic states including the self-cleaning process are tested by three cycles with almost no structural color degeneration is observed as shown in **Figure** 6. Recently, $TiO_2$ based dielectric metasurface gained lots of attention, with itself as a good photocatalyst and can be used to induce wettability transition under UV light irradiation; metasurface constructed by similar functional materials could provide far more exciting applications in the near future.[7, 23, 25, 38, 46] Here, the switching of functionality signifies that we can choose the required state for the metasurface as demanded, i.e., ODP modified superhydrophobic state can be used to store the sample once the structures are fabricated, this can be used to protect or easily clean them during daily use. Then, the superhydrophilic state can be induced once a specific application like dynamic response or perfect solution matching are needed.

With two samples and six metasurface patterns on each (both with wettability supporting structures), we check more details of the remaining particles on the surface by SEM and summarize the repeating cleaning results in *Supporting Information Section 8* (one cycle includes both self-cleaning and color response). On both samples, at least three metasurface fields (half of the total number) exist with less than five particles after three repeating rounds, and most of the fields remain clean with only one field that still has more than twenty particles. These results further demonstrate that the self-cleaning process works for the sample with wettability supporting design, as all the six metasurfaces are not cleaned at the first cycle for the sample without wettability supporting structure (*Supporting Information Section 7*).

## 3. Conclusion

By using a structural color metasurface as a platform, we show a strightforward route to achieve versatile functionality with both self-cleaning and dynamic color response based on the superwettability property. The self-cleaning ability is achieved at the ODP modified quasi-superhydrophobic state, which can be used to efficiently self-clean particle-based contamination with water. The superhydrophilic state response is used for obtaining a rapid dynamic structural color change. We found that the superhydrophilic state is stably once the surface is treated with $O_2$ plasma. Furthermore, the proposed metasurface can easily be switched between these two states, no performance degradation is observed after three repeating cycles.

As a well-developed topic than optical metasurface, wettability design has many more available methods beyond that we present here. For example, the two-step fabrication process or uniform wettability supporting structures can be replaced by other proper strategies to reduce the complexity in fabrication or improve the device's performance. With properly design and wettability states, an extension to metasurfaces will provide more functionalities except self-cleaning and color response in this work.[37] Furthermore, by constructing metasurfaces with materials like ZnO or TiO$_2$ that are widely used in wettability design and transition, these functionalities may be easily incorporated into the metasurfaces.[46] Our design strategy can also be extended to the fabrication of other optical metasurfaces, which can be used for applications in more complex environments.

## 4. Experimental methods

*Fabrication of the metasurface*

The metasurface with wettability supporting structures is fabricated by the following steps. First, a Cr mask corresponding to the designed QR code image is fabricated on an ITO substrate with 700 nm a-Si by standard EBL process consisting of electron beam resist spin coating, structure patterning, developing of the exposed resist, Cr mask (20 nm) deposition and lift off. Then, by using the Cr mask on the sample, we spin coat another layer of UV sensitive negative resist and perform UV laser lithography with high throughput (around 10 mins to pattern the 10 x 10 mm$^2$ area). Developing the resist after UV laser lithography results in two masks (Cr and resist mask). Further, the etching step provides both nano-sized metasurface and wettability supporting structures together. Finally, the sample after etching is cleaned in acetone and Cr etching solution separately to remove the remaining resist and Cr mask.

For the hydrophobic treatment, isopropanol is used as a solvent to produce the Octadecylphosphonic acid (ODP, Sigma-Aldrich) solution. Before modifying the sample with ODP, the sample is treated by O$_2$ plasma for 5 mins to increase the polarity of the structure, then it is immersed in the ODP solution for 1 h. The sample is then rinsed by isopropanol and blown with nitrogen gas. To achieve the superhydrophilic sate, the same sample is treated with O$_2$ plasma for 5 mins, and all the measurements are finished within 2 hours.

*Wettability measurement*

Wettability of the sample is evaluated by measuring the static value using a CA meter in ambient environment. A 5 μL droplet of deionized water is used in the measurement. The reported CA value in the main text is an averaged value over three measurements.

*Optical measurement*

The reflective spectrum is measured with a spectrometer connected to an optical microscope. By using a beam splitter, the spectrum of the metasurface is collected with a fiber collimator and analyzed by the spectrometer, while the color image is recorded by the same setup at the other port with a camera. The reflective spectrum in water is recorded by immersing the sample in a glass channel that is filled with water. The QR code color image is recorded with a microscope (objective lens: 20x, NA 0.45) illuminated by white light. The dynamic response in the video is from a commercial optical microscope from the top-view mode, which can be used to record the image with water droplet on the surface (the image qualities are different for both setups as shown in *Supporting Information Section 9*).

*Self-cleaning and physical stability measurement*

The artificial contamination consists of carbon particles with nano to micro sizes, they are spread on the whole surface before cleaning. Then the carbon particles are washed with 1 ml of water to record the cleaning video at a title angle around 45°. To record the optical image and check with SEM, the sample is further rinsed with water gently and blow slightly with nitrogen gas for drying. For the repeatability testing, the sample after self-cleaning examination is treated by $O_2$ plasma for another 5 mins to induce the superhydrophilic color change state, then it is modified in ODP solution for the following testing once the image in water is recorded. The process is repeated for three times for testing.

**Acknowledgments**

This project has received funding from the European Research Council (ERC) under the European Union's Horizon 2020 research and innovation program (grant agreement No. 724306) and the Deutsche Forschungsgemeinschaft (DFG No. ZE953/11-1).